\documentclass[%
 reprint,
superscriptaddress,
nofootinbib,
 amsmath,amssymb,
 aps,
 prd,
]{revtex4-2}

\usepackage{graphicx}
\usepackage{dcolumn}
\usepackage{bm}
\usepackage{xcolor}
\usepackage{orcidlink}
\usepackage{longtable}

\DeclareRobustCommand{\VAN}[3]{#2}
\let\VANthebibliography\thebibliography
\def\thebibliography{\DeclareRobustCommand{\VAN}[3]{##3}\VANthebibliography}

\newcommand{\msun}{{\,\rm M_\odot}}

\newcommand{\cpm}{\,{\rm cm}^2\,{\rm g}^{-1}}

\def\jcap{J. Cosmol.  Astropart. Phys.}

\def\apj{ApJ}

\def\apjl{ApJ}
\def\mnras{MNRAS}
\def\araa{ARA\&A}

\def\physrep{Phys. Rep.}
\def\nat{Nature}

\def\prd{Phys. Rev. D}

\begin{document}

\preprint{APS/123-QED}

\title{Novel density profile for isothermal cores of dark matter halos}

\author{Vinh Tran\,\orcidlink{0009-0003-6068-6921}}
 \affiliation{Department of Physics and Kavli Institute for Astrophysics and Space Research, Massachusetts Institute of Technology, Cambridge, MA 02139, USA}
 \email{vinhtran@mit.edu}

\author{Xuejian Shen\,\orcidlink{0000-0002-6196-823X}}
\affiliation{Department of Physics and Kavli Institute for Astrophysics and Space Research, Massachusetts Institute of Technology, Cambridge, MA 02139, USA}

\author{Mark Vogelsberger\,\orcidlink{0000-0001-8593-7692}}
\affiliation{Department of Physics and Kavli Institute for Astrophysics and Space Research, Massachusetts Institute of Technology, Cambridge, MA 02139, USA}

\author{Daniel Gilman\,\orcidlink{0000-0002-5116-7287}}
\affiliation{Department of Astronomy \& Astrophysics, University of Chicago, Chicago, IL 60637, USA}

\author{Stephanie O'Neil\,\orcidlink{0000-0002-7968-2088}}
\affiliation{Department of Physics \& Astronomy, University of Pennsylvania, Philadelphia, PA 19104, USA}
\affiliation{Department of Physics, Princeton University, Princeton, NJ 08544, USA}

\author{Cian Roche\,\orcidlink{0000-0002-3400-6991}}
\affiliation{Department of Physics and Kavli Institute for Astrophysics and Space Research, Massachusetts Institute of Technology, Cambridge, MA 02139, USA}

\author{Oliver Zier\,\orcidlink{0000-0003-1811-8915}}
\affiliation{Center for Astrophysics, Harvard \& Smithsonian, 60 Garden St, Cambridge, MA 02138, USA}

\author{Jiarun Gao}
\affiliation{University of Rochester, 500 Joseph C Wilson Blvd, Rochester, NY 14627, USA}

\date{\today}

\begin{abstract}
    We present a novel analytic density profile for halos in self-interacting dark matter (SIDM) models, which accurately captures the isothermal-core configuration, i.e. where both the density and velocity dispersion profiles exhibit central plateaus in the halo innermost region. Importantly, the profile retains a simple and tractable functional form. We demonstrate analytically how our density profile satisfies the aforementioned conditions, with comparisons to other contemporary functional choices. We further validate the profile using idealized N-body simulations, showing that it provides excellent representations of both the density and velocity dispersion profiles across a broad range of evolutionary stages, from the early thermalization phase to the late core-collapse regime. As a result of its accuracy and simplicity, the proposed profile offers a robust framework for analyzing halo evolution in a variety of SIDM scenarios. It also holds practical utility in reducing simulation needs and in generating initial conditions for simulations targeting the deep core-collapse regime.
\end{abstract}

\maketitle


\section{Introduction}
\label{sec:intro}

Despite the success of the $\Lambda$CDM (cosmological constant $\Lambda$ plus collisionless cold dark matter, CDM) model in explaining the large-scale structure of the universe~\citep[e.g.][]{Blumenthal1984,Davis1985}, challenges remain in matching simulations to observations of small-scale structures~\citep[e.g.][]{Bullock2017,Sales2022}. For example, many dwarf galaxies exhibit kpc-size cores in contrast to the universal cuspy density profiles seen in CDM-only simulations (core-cusp problem)~\citep[e.g.][]{Flores1994,Moore1994}. The dwarf spheroidal galaxies in the Local Group are found to be systematically less dense compared to the subhalos in CDM-only simulations (the too-big-to-fail problem)~\citep[e.g.][]{BoylanKolchin2011}. Although stellar feedback processes have been shown to alleviate these
tensions~\citep[e.g.][]{Governato2010,Pontzen2012}, a population of compact dwarf galaxies is missing in simulations of CDM (plus baryons) that can produce DM cores~\citep[e.g.][]{Santos-Santos2020}. The diversity problem of dwarf galaxy structures is still a challenge for the CDM framework~\citep[e.g.][]{Oman2015,Kaplinghat2019}.

Self-interacting dark matter (SIDM) is an important class of DM models that serves as an alternative to CDM and is well motivated by hidden sector models beyond the standard model~\citep[e.g.][]{Ackerman2009, Arkani-Hamed2009, Cyr-Racine2013}. SIDM provides promising solutions to many small-scale problems (see \cite{Tulin2018} for a review). In most cases, DM self-interactions are considered elastic for simplicity and will be the main focus of this paper, although inelastic DM self-interactions have raised some attention recently~\citep[e.g.][]{Essig2019,Vogelsberger2019,Shen2021,Shen2022,O'Neil2023}. The elastic collisions between DM particles result in the thermalization of the inner part of the halo and lead to the formation of cores~\citep[e.g.][]{Vogelsberger2012,Rocha2013,Elbert2015}, which could explain the core-cusp and the too-big-to-fail problems. Meanwhile, the sensitivity of SIDM halo structure to halo concentration and baryon dominance results in more diverse rotation curves compared to the CDM case and is more aligned with observations~\citep[e.g.][]{Oman2015,Kaplinghat2019,Jiang2023}. In recent years, SIDM models with very large self-interaction cross sections at lower velocity ranges have gained popularity. It has implications for the diversity problem of dwarf galaxy structures and the dense compact substructures of galaxy clusters found in strong gravitational lensing observations~\citep[e.g.][]{Gilman2023}. In this regime, the efficient heat conduction and the negative heat capacity of the SIDM halo result in gravothermal collapse \citep[e.g.][]{Burkert2000,Kochanek2000,Balberg2002}, originally discussed in the context of globular clusters, e.g. \cite{Lynden-Bell+Eggleton1980}.

Following the initial thermalization phase and throughout the gravothermal collapse process, until reaching the short-mean-free-path regime when heat conduction is only effective at the surface of the core, the halo core exhibits self-similarity in both the density and velocity dispersion profiles. That is, throughout its evolution, the halo profiles can be characterized by specific values of scale density, radius, and velocity dispersion. This behavior has been well described theoretically~\citep[e.g.,][]{Balberg2002} and confirmed in simulations for both velocity-independent and velocity-dependent self-interacting cross sections~\citep[e.g.,][]{Outmezguine2023, Yang2023}. In addition, when these scale quantities are normalized by their initial values (or by the corresponding NFW values), and time is rescaled by a collapse timescale (inversely proportional to the self-interaction cross section), the evolution of halos with different parameters would become almost identical. Such universality, discussed and observed in~\citep{Zhong2023}, as well as the aforementioned self-similar natures, provide the foundation for adopting particular functional forms to describe the density structure of halos. In this self-similar regime, the halo core develops characteristic plateaus in both the density and velocity dispersion profiles. The emergence of a central plateau in the density profile, commonly referred to as the ``core configuration'', is a key feature of SIDM models and is instrumental in addressing the core-cusp problem, as discussed above. In this context, we also define the ``isothermal-core configuration'' as a state in which both the density and velocity dispersion remain approximately constant within the central region of the halo. Such behaviors of SIDM have been found in cosmological and isolated N-body simulations~\citep[e.g.][]{Colin2002,Vogelsberger2012,Rocha2013,Elbert2015,Yang2023}. In the outer regions of the halo, however, the number density and velocity dispersion of DM particles are significantly smaller than in the core, resulting in negligible scattering interactions over the age of the Universe. As a consequence, the density profile asymptotically approaches the NFW profile~\citep{NFW1996, NFW1997} commonly observed in CDM.

To characterize the properties of DM halos in simulations, it is common to fit the density profiles with certain functional choices based on the core configuration~\citep[e.g.][]{Read2016,Robertson2017,Yang2023,Fischer2024,Yang2024,Palubski2024}. Efforts to derive analytical density profiles based on the isothermal-core configuration have also been made, either by enforcing the isothermal condition in the Jeans equation~\citep[e.g.][]{Kaplinghat2014,Robertson2021,Jiang2023} or within the more complex gravothermal fluid formalism~\citep[e.g.][]{Yang2023}. However, despite the advantage of directly capturing the isothermal core and being robust in accommodating baryonic configurations, these profiles lack a closed-form functional expression. Here, we introduce a novel functional form for the density profile that provides a close approximation to the isothermal-core configuration. Section \ref{sec:profiles} details the different contemporary density profiles, as well as the motivation and derivation of our profile. We also examine the analytical behavior of each density profile and the corresponding velocity dispersion profiles. Section \ref{sec:simulation_comparison} inspects the fitting and reconstructing results of the profiles in the context of isolated N-body simulations. We conclude the investigation with a summary in Section \ref{sec:conclusions} and discuss potential further applications of the profiles in future SIDM studies.


\begin{table*}
    \centering
    \addtolength{\tabcolsep}{10pt}
    \def\arraystretch{2.5}
    \begin{tabular}{c c c c c}
        \hline
        Density Profile & $N_{\rm{DOF}}$ & Functional Form & $\rho(r_{\rm{c}} < r < r_{\rm{s}}^{\prime})$ & $r_{\rho/2} (r_{\rm{c}} \ll r_{\rm{s}}^{\prime})$ \\ [1ex]
        \hline\hline

        $\rho_{\rm{R16}}$ & 3 & $\rho_{\rm{c}} \frac{2}{3\left(r/r_{\rm{s}}^{\prime}\right)^2} \frac{\partial}{\partial r/r_{\rm{c}}} \Big[\tanh{r/r_{\rm{c}}} \left(\ln{r/r_{\rm{s}}^{\prime}} - \frac{r/r_{\rm{s}}^{\prime}}{1+r/r_{\rm{s}}^{\prime}}\right)\Big]$ & $\rho \propto r^{-1}$ & $r_{\rm{c}} \tanh^{-1}{0.75}$ \\ [1ex]

        $\rho_{\rm{R17-F24}}$ & 4 & $\rho_{\rm{c}} \Big[ \left(1 + \left(r/r_{\rm{c}}\right)^4\right)^{n/4} \left(1+r/r_{\rm{s}}^{\prime}\right)^{3-n} \Big]^{-1}$ & $\rho \propto r^{-n}$ & $r_{\rm{c}} \left(2^{4/n}-1\right)^{1/4}$ \\ [1ex]

        $\rho_{\rm{Y23}}$ & 4 & $\rho_{\rm{c}} \Big[ 1 + \left(r/r_{\rm{c}}\right)^n \left(1 + r/r_{\rm{s}}^{\prime}\right)^{3-n} \Big]^{-1}$ & $\rho \propto r^{-n}$ & $r_{\rm{c}}$ \\ [1ex]

        $\rho_{\rm{T25}}$ & 4 & $\rho_{\rm{c}} \left( \frac{\tanh{r/r_{\rm{c}}}}{r/r_{\rm{c}}} \right)^n \Big[ \left(1 + \left(r/r_{\rm{s}}^{\prime}\right)^2\right)^{\left(3-n\right)/2} \Big]^{-1}$ & $\rho \propto r^{-n}$ & $r_{\rm{c}} f^{-1}(0.5^{1/n})$ \\ [1ex]
        \hline
    \end{tabular}
    \caption{Summary of analytical density profiles. The profile parameters (detailed further in the text) $\beta = 4$, and $\gamma = 2$ are fixed for $\rho_{\rm{R16}}$, $\rho_{\rm{R17-F24}}$, and $\rho_{\rm{T25}}$, respectively. $N_{\rm{DOF}}$ is the number of DOF in the different profiles. $\rho(r_{\rm{c}} < r < r_{\rm{s}}^{\prime})$ and $r_{\rho/2} (r_{\rm{c}} \ll r_{\rm{s}}^{\prime})$ show the behaviors of the density profiles in the intermediate region and the approximated core half-density radii, respectively. For our density profile ($\rho_{\rm{T25}}$), we define $f(x) = \tanh{x}/x$.}
    \label{tab:density_profile_summary}    
\end{table*}

\section{Analytical Density Profiles}
\label{sec:profiles}

\subsection{Core density profiles}
\label{ssec:core_profiles}

In this section, we will review several existing analytical density profiles proposed for a cored DM distribution. The physical driver for coring is usually at the center of the DM halo, so these profiles should all transition to the NFW profile at large radii. The NFW profile represents a universal scale-invariant stable configuration of halos in the CDM model observed in cosmological N-body simulations, following
\begin{equation}
    \label{eqn:nfw_density_profile}
    \rho_{\rm{NFW}} (r) = \frac{\rho_s}{\left(r/r_{\rm{s}}\right) \left(1 + r/r_{\rm{s}}\right)^2},
\end{equation}
where, $\rho_{\rm{s}}$ and $r_{\rm{s}}$ are the scale density and radius. Here, $r$ is the spherical radial coordinate. This profile exhibits a cuspy central density, with steep divergence as $r$ approaches 0. To instead describe the core configuration, a common approach is to modify the NFW profile with a function that flattens the density at small radii but maintains convergence to the NFW profile at large radii. A density profile that utilizes one such approach is the Robertson-Fischer $\rho_{\rm{R17-F24}}$ profile~\citep{Robertson2017,Fischer2024}
\begin{equation}
    \label{eqn:Robertson-Fischer_density_profile_NFW}
    \rho^{\ast}_{\rm{R17-F24}} (r) = \frac{\rho_{\rm{c}}}{\left(1 + \left(r/r_{\rm{c}}\right)^\beta\right)^{1/\beta} \left(1+r/r_{\rm{s}}^{\prime}\right)^2},
\end{equation}
with $\beta \in \left[2;4\right]$\footnote{Similar functional forms are also utilized in \cite{Yang2024,Palubski2024} among others.}. This power-law, often fixed and taking the value of $\beta = 4$, results in the density profile that converges almost perfectly to the NFW profile for all radii larger than the core radius. $\rho_{\rm{c}}$ and $r_{\rm{c}}$ represent the characteristic core density and radius, respectively. $r_{\rm{s}}^\prime$ is a scale radius, characterizing the NFW tail. The prime symbol is used to distinguish this parameter from the initial NFW scale radius $r_{\rm{s}}$. Although $r_{\rm{s}}^\prime$ is similar in nature, it should not be confused with $r_{\rm{s}}^\prime$.

As an alternative to $\rho_{\rm R17-F24}$, the Read profile $\rho_{\rm{R16}}$ approximates the enclosed mass profiles of cored halos as
\begin{equation}
    \label{eqn:Read_mass_profile}
    M_{\rm{R16}}(r) = M_{\rm{NFW}}^{\prime}(r) \tanh{\left(r/r_{\rm{c}}\right)},
\end{equation}
where $M_{\rm{NFW}}^{\prime}\left(r\right)$ is the NFW mass profile derived from the corresponding NFW scale density $\rho_{\rm{s}}^{\prime}$ and scale radius $r_{\rm{s}}^{\prime}$\footnote{Originally, the Read profile was used to characterize the baryon-induced flat core of DM halos, with an additional exponent $\alpha$ in the $\tanh^\alpha{\left(r/r_{\rm{c}}\right)}$ term, controlling the core flatness. However, it has also been used in the context of SIDM as a convenient fitting function~\citep[e.g.][]{Jiang2023,Yang2023,Yang2024}.}. The characteristic core density $\rho_{\rm{c}}$ of the Read profile can be recovered as $\rho_{\rm{c}} = 1.5 \rho_{\rm{s}}^{\prime} r_{\rm{s}}^{\prime} / r_{\rm{c}}$. It must be noted here that the scale density and radius of the Read and other profiles (indicated by the prime notation to be distinguished from the NFW profile parameters) are not necessarily the same as the initial NFW profile parameters, with their values changing throughout the evolution of the halo.

For the Robertson-Fischer and Read profiles, when the condition of $r_{\rm{c}} \ll r_{\rm{s}}^\prime$ is satisfied, the density profile asymptotically approach a power-law of $\rho \sim r^{-n}$ in the intermediate region of $r_{\rm{c}} \ll r \ll r_{\rm{s}}^{\prime}$. Here, $n$ is the transition index, taking the value of $n=1$. This index controls the steepness of the transition between the constant-density core ($\rho \sim \text{const}$) and the NFW tail ($\rho \sim r^{-3}$), even in the case of $r_{\rm{c}} \sim r_{\rm{s}}^\prime$, where intermediate region becomes ill-defined. It has been demonstrated in \cite{Balberg2002} that in the gravothermal fluid formalism, a self-similar solution for SIDM halos can be found in which $n=2.19$, provided the collisional mean free path is much larger than the gravitational scale height. This is very similar to the value of $n \simeq 2.5$ observed in simulations (as observed in Section \ref{ssec:fitting_results}). However, we find that this is a difficult constraint. These values, which can also vary throughout the halo evolution, as detailed later in Section \ref{ssec:fitting_results}, are significantly different from the value of $n=1$ taken by the Read and Robertson-Fischer approaches. Thus, modifications or alternative forms of density profiles are necessary. Accommodating this variation requires the Robertson-Fischer profile to increase the number of degrees of freedom (DOF), taking the form
\begin{equation}
    \label{eqn:Robertson-Fischer_density_profile}
    \rho_{\rm{R17-F24}} (r) = \frac{\rho_{\rm{c}}}{\left(1 + \left(r/r_{\rm{c}}\right)^\beta\right)^{n/\beta} \left(1+r/r_{\rm{s}}\right)^{3-n}}.
\end{equation}
On the other hand, we keep the functional form of the Read profile, since the initial construction of it involves the NFW mass profile and is more difficult to modify. 

Another straightforward approach to approximating the core configuration is the Yang profile $\rho_{\rm{Y23}}$~\citep{Yang2023}, which utilizes a triple power-law to represent the dark matter halo structure,
\begin{equation}
    \label{eqn:Yang_density_profile}
    \rho_{\rm{Y23}} (r) = \frac{\rho_{\rm{c}}}{1 + \left(r/r_{\rm{c}}\right)^n \left(1 + r/r_{\rm{s}}\right)^{3-n}},
\end{equation}
where $n$ is originally set to $n=2.19$. However, here we allow $n$ to vary as a free parameter.

The majority of the evolution of an SIDM halo stays in the long mean free path (LMFP) limit, where the mean free path of SIDM is much larger than the gravitational scale height of the system. A self-similar solution of the density profile has been found in this regime~\citep[e.g.][]{Lynden-Bell+Eggleton1980, Balberg2002}. One indication of such behavior is the similarity in transitional behavior across different epochs, as evidenced by the stability of $n$ throughout the halo evolution. This suggests that measuring $n$ is more informative than assuming fixed values when probing self-similarity.

Additionally, we note that while the core density $\rho_{\rm{c}}$ retains a consistent physical meaning across the Read, Robertson-Fischer, and Yang profiles, the interpretations of the core radius $r_{\rm{c}}$ differ. This is a result of the different functional forms of the profiles. To facilitate a more meaningful comparison of core sizes across different profiles, we introduce a new parameter $r_{\rho/2}$, the core half-density radius, defined as the radius at which the halo density falls to half the value of $\rho_{\rm{c}}$. Typically, $r_{\rho/2}$ scales linearly with $r_{\rm{c}}$, with the proportionality factor determined by the value of $n$ (as well as $r_{\rm{s}}^{\prime}$ when $r_{\rm{c}} \sim r_{\rm{s}}^{\prime}$). A summary of the mapping between $r_{\rm{c}}$ and $r_{\rho/2}$ in the $r_{\rm{c}} \ll r_{\rm{s}}^{\prime}$ regime is presented in Table \ref{tab:density_profile_summary}.

\subsection{Isothermal-core density profile}
\label{ssec:isothermal-core_profile}

Although the aforementioned profiles capture the cored mass distribution of DM halos, they were not designed to reproduce the kinematic structure of SIDM halos. The velocity dispersion profiles they produce often deviate significantly from simulations and the isothermal-core configuration, which requires a constant velocity dispersion in the core.

To construct a density profile that satisfies the isothermal-core configuration, we start from the Jeans equation~\citep[e.g.][]{Kaplinghat2014,Robertson2021,Jiang2023}
\begin{align}
    \label{eqn:Jeans_equation}
    \frac{\partial}{\partial r} \Big[\rho(r) \sigma_{\rm{1D}}^2(r)\Big] &= - \rho(r) \frac{\partial \Phi(r)}{\partial r},
\end{align}
where we assume spherical symmetry and isotropy. Considering only dark matter, the gravitational potential term can be written as $\partial \Phi(r) / \partial r = G M(r) / r^2$, with $M(r)$ as the enclosed mass profile, obtained from $\partial M(r) / \partial r = 4 \pi r^2 \rho(r)$.\footnote{More complicated systems involving both baryonic and dark matter would require treatments utilizing Poisson's equation.} The one-dimensional velocity dispersion $\sigma_{\rm{1D}}$ is taken to be constant within a specified radius, beyond which the profile transitions to an NFW form. This method is relatively accurate and can accommodate adiabatic contractions caused by baryons. However, no closed-form expression exists for the density profile calculated from this approach, as it necessitates numerical calculations and a more complicated boundary-stitching procedure~\citep{Jiang2023}.


As our goal is to arrive at an analytical density profile that closely satisfies the isothermal-core configuration while maintaining a well-defined functional form, we take a similar but slightly more simplistic approach. First and foremost, we require our density profile to exhibit similar behaviors to the profiles detailed in Section \ref{ssec:core_profiles}, that is: having a radius-independent density at lower radii, a transition index of $n$ in the intermediate region, and a power-law of $r^{-3}$ similar to the NFW profile at the halo edge. In agreement with simulation results, we assume a constant core velocity dispersion $\sigma_{\rm{c}}$ within the region of $r \lesssim r_{\rm{c}}$, leading to the following dimensionless set of equations
\begin{align}
    \label{eqn:unit-free_mass_equation}
    \frac{\partial \bar{M} (\bar{r})}{\partial \bar{r}} &= \hspace{0.1cm} \bar{r}^2 \bar{\rho} (\bar{r}), \\
    \label{eqn:unit-free_Jeans_equation}
    \frac{\partial \bar{\rho} (\bar{r})}{\partial \bar{r}} &= - \frac{1}{I} \frac{\bar{\rho} (\bar{r}) \bar{M} (\bar{r})}{\bar{r}^2},
\end{align}
with, $\bar{r} = r / r_{\rm{c}}$, $\bar{\rho} (\bar{r}) = \rho (r) / \rho_{\rm{c}}$, and $\rho$ being our desired density profile. Here,
$\bar{M} (\bar{r}) = M (r) / 4 \pi r_{\rm{c}}^3 \rho_{\rm{c}}$ and $I = \sigma_{\rm{c}}^2 / 4\pi G \rho_{\rm{c}} r_{\rm{c}}^2$. This parameter $I$ can be interpreted as a characteristic core kinetic-to-gravitational (K–G) ratio under the isothermal-core assumption, which can be calculated from $\bar{\rho}$ as
\begin{equation}
    \label{eqn:isothermal_K-G_ratio}
    I = \frac{- \bar{\rho}^3 \bar{r}}{2 \bar{\rho}\,\partial \bar{\rho} / \partial \bar{r} - \bar{r} \left(\partial \bar{\rho} / \partial \bar{r}\right)^2 + \bar{r} \bar{\rho}\,\partial^2 \bar{\rho} / \partial \bar{r}^2}.
\end{equation}
For the assumption of a constant core velocity dispersion $\sigma_{\rm{c}}$ to hold, the parameter $I$ must remain independent of radius. Here, we make two additional assumptions. First, we assume that $r_{\rm{c}} \ll r_{\rm{s}}^{\prime}$, which allows us to simplify the functional form to a constant-density core transitioning to a slope of $n$ at larger radii. Second, we assume that if $I \sim \text{const}$ up to order $\mathcal{O}(\bar{r}^2)$, then $I$ remains approximately radius-independent throughout the majority of the halo core. As detailed in Appendix \ref{apd:sigma_c_T25}, this is satisfied by the functional form of
\begin{equation}
    \label{eqn:TVS_density_profile_simple}
    \rho(r) = \rho_{\rm{c}} \left( \frac{\tanh{r/r_{\rm{c}}}}{r/r_{\rm{c}}} \right)^n.
\end{equation}
Including the necessary modifications to satisfy the density structure requirement of the NFW tail, we introduce our density profile $\rho_{\rm{T25}}$\footnote{ We note that our profile closely resembles the one employed in~\citep{Roberts2024}, as both adopt the $\tanh{r} / r$ functional approach.}
\begin{equation}
    \label{eqn:TVS_density_profile}
    \rho_{\rm{T25}} (r) = \rho_{\rm{c}} \left( \frac{\tanh{r/r_{\rm{c}}}}{r/r_{\rm{c}}} \right)^n \frac{1}{\left(1 + \left(r/r_{\rm{s}}\right)^\gamma\right)^{\left(3-n\right)/\gamma}}.
\end{equation}
Here, $\gamma$ functions similarly to the nuisance parameter $\beta$ in the Robertson-Fischer profile, in that they both control the sharpness of the core. We find that the best results, i.e., where the velocity dispersion of the halo core least depends on the radius, are provided by $\gamma \geq 2$ with the detailed explanation presented in Appendix \ref{apd:sigma_c_T25}. We fix $\gamma = 2$ for the remainder of our analysis. Similar adjustments can also be made for other profiles. However, we find that such modifications have no significant effect on the quality of the profile fits to simulations, nor on the agreement between the reconstructed velocity dispersion profiles and the desired isothermal-core configuration. 

From Equation \ref{eqn:TVS_isothermal_K-G_ratio_gamma_2} and the definition of $I$, we also retrieve the core velocity dispersion of our profile
\begin{equation}
    \label{eqn:TVS_sigma_c}
    \tilde{\sigma}_{\rm{c,T25}} = \sqrt{\frac{4 \pi G \rho_{\rm{c}} r^2_{\rm{c}}} {2n + 3\left(3-n\right) \left(r_{\rm{c}}/r_{\rm{s}}^{\prime}\right)^2} }.
\end{equation}
The tilde denotes that the expression serves as an approximation; a more accurate value can be obtained by averaging the reconstructed velocity dispersion profile within the core radius. Nevertheless, the deviation between the two approaches is typically of the order of $1\%$. Typically, $n$ is close to 3 and $r_{\rm{c}} \ll r_{\rm{s}}^{\prime}$, causing the second term in the denominator to become negligible. For $n \simeq 2.5$, the core half-density radius also satisfies $r_{\rho/2} \simeq r_{\rm{c}}$. Table \ref{tab:density_profile_summary} summarizes the characteristics of the analytical density profiles under inspection.

\subsection{Analytical characteristics of density profiles}
\label{ssec:profiles_analysis}

\begin{figure*}
    \centering
    \includegraphics[width=0.99 \textwidth]{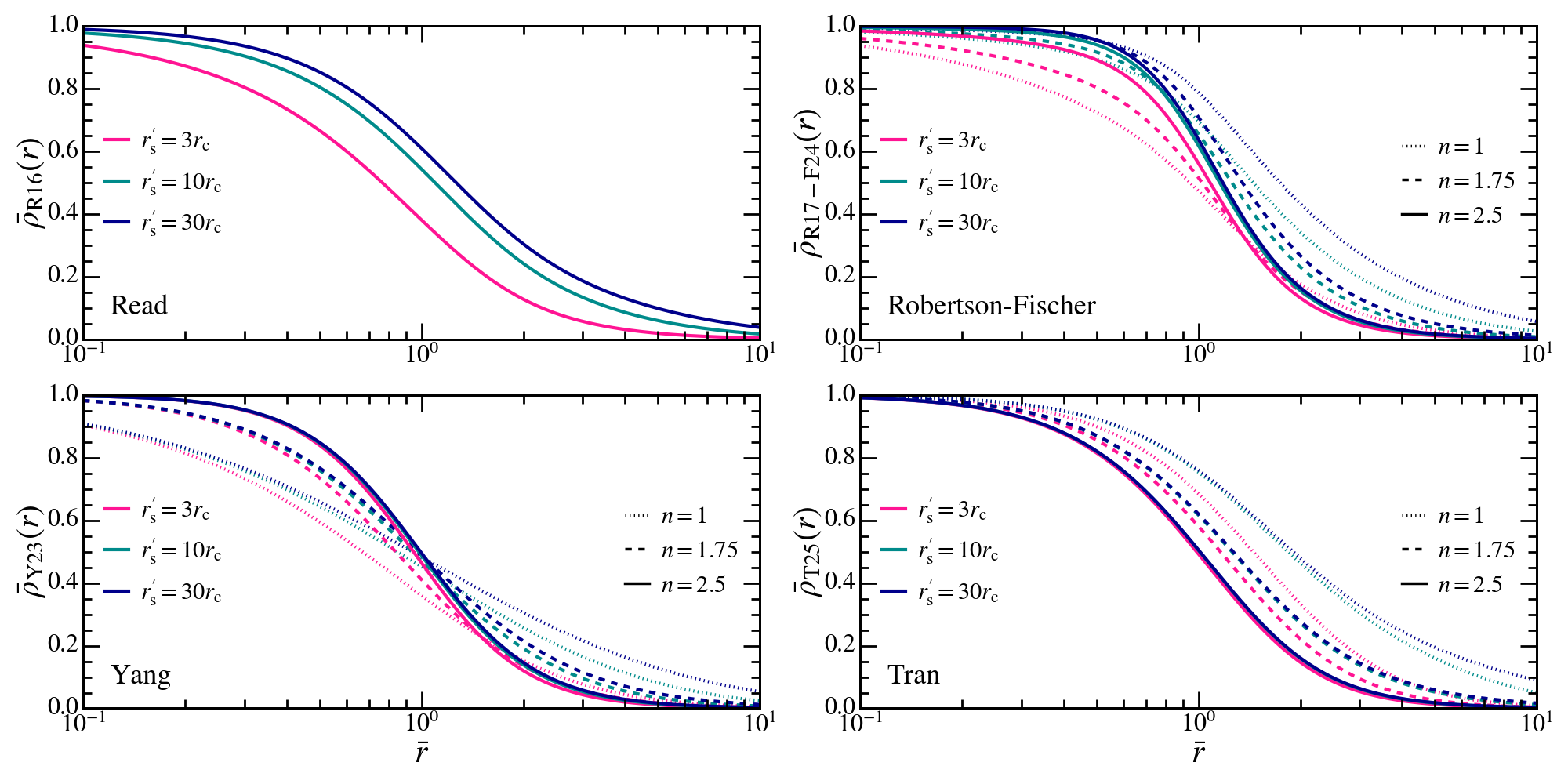}
    \caption{The density profiles of the Read (top left), Robertson-Fischer (top right), Yang (bottom left), and our (bottom right) profiles. Densities and radii are scaled with the characteristics core density $\rho_{\rm{c}}$ and core radius $r_{\rm{c}}$. Each profile family is shown with three choices of $r_{\rm{s}} / r_{\rm{c}} = 3, 10, 30$ (in pink, green, and blue) and three choices of $n = 1, 1.75, 2.5$ (in dotted, dashed, and solid lines). The choice of $n$ is not relevant in the context of the Read profile.}
    \label{fig:density_profiles}
\end{figure*}

\begin{figure*}
    \centering
    \includegraphics[width=0.99 \textwidth]{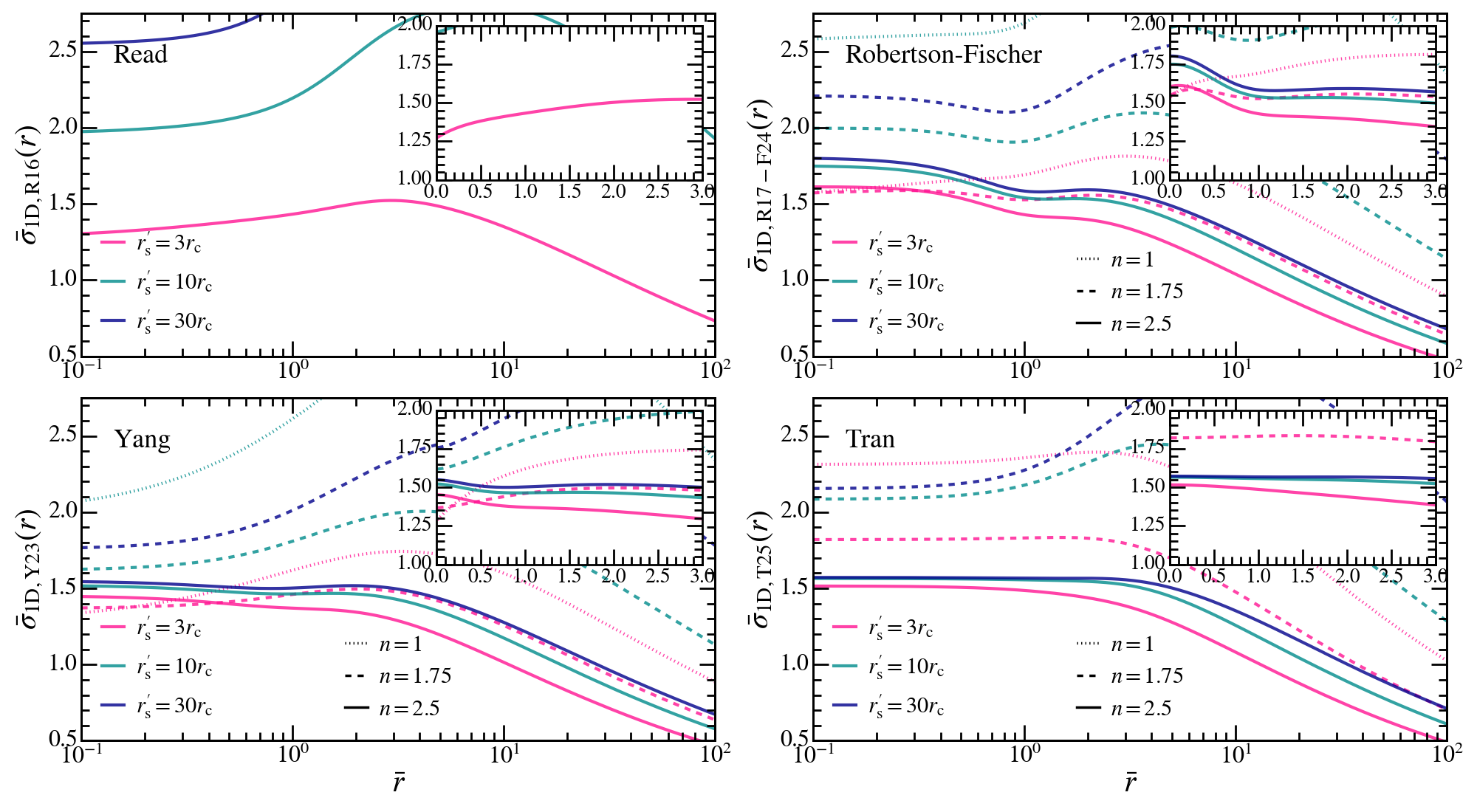}
    \caption{The velocity dispersion profiles reconstructed from the density profiles presented in Figure \ref{fig:density_profiles} using the spherically symmetric Jeans equation (Equation \ref{eqn:Jeans_equation}). Velocity dispersions and radii are scaled with $\sqrt{G \, \rho_{\rm{c}} \, r_{\rm{c}}^2}$ and $r_{\rm{c}}$, respectively. Again, the Read (top left), Robertson-Fischer (top right), Yang (bottom left), and our (bottom right) profiles are shown with three choices of $r_{\rm{s}} / r_{\rm{c}} = 3, 10, 30$ (in pink, green, and blue), as well as three choices of $n = 1, 1.75, 2.5$ (in dotted, dashed, and solid lines) for the relevant profiles. The sub-panels present a zoomed-in view of the velocity dispersion profiles within the expected isothermal region of the core $r \lesssim 3 \, r_{\rm{c}}$.}
    \label{fig:velocity_dispersion_profiles}
\end{figure*}

Figure \ref{fig:density_profiles} shows the Read density profile ($\rho_{\rm{R16}}$), the modified Robertson-Fischer profile (hereafter referred to as the Robertson-Fischer profile; $\rho_{\rm{R17-F24}}$), the Yang profile ($\rho_{\rm{Y23}}$), and the profile proposed and studied in this work ($\rho_{\rm{T25}}$). Densities are normalized as $\bar{\rho} = \rho / \rho_{\rm{c}}$ and shown as functions of the normalized radius $\bar{r} = r / r_{\rm{c}}$. We show density profiles with three choices of edge-core ratios, $r_{\rm{s}}^{\prime} / r_{\rm{c}} = 3, 10, 30$, and three choices of the transition index, $n = 1, 1.75, 2.5$. As expected, all profiles show similar morphology by construction, except for the Read profile being limited by its DOFs. In both the Yang profile and our profile, we find that the same values of $n$ produce similar density profiles despite the variation of $r_{\rm{s}}^{\prime}$. This persists even in the regime where $r_{\rm{c}} \sim r_{\rm{s}}^{\prime}$ for sufficiently high values of $n \sim 2.5$, indicating a dominant role of $n$ in determining the transition between the halo core and the outer edge. Such a behavior is not observed in the Robertson–Fischer profile, even for the $r_{\rm{c}} \ll r_{\rm{s}}^{\prime}$ regime.

Figure~\ref{fig:velocity_dispersion_profiles} shows the normalized one-dimensional velocity dispersion profiles $\bar{\sigma}_{\rm{1D}} = \sigma_{\rm{1D}} / \sqrt{G \bar{\rho}_{\rm{c}} \bar{r}_{\rm{c}}^2}$ reconstructed from Figure~\ref{fig:density_profiles}'s density profiles using the spherically symmetric Jeans equation (Equation~\ref{eqn:Jeans_equation}). The sub-panels display the velocity dispersion profiles in linear scale, with a zoomed-in focus on the $r \lesssim 3\,r_{\rm{c}}$ region, where the profile is expected to exhibit isothermal behavior. For the Robertson-Fischer and Yang profiles, the velocity dispersion profiles with higher values of $n$ tend to approach the isothermal-core configuration in the innermost region of $r \ll r_{\rm{c}}$, albeit with relatively abrupt core transitions. In contrast, our profile exhibits smoother transitions across parameter choices, with the velocity dispersion remaining nearly constant throughout the core, indicating a closer approximation to the isothermal-core configuration.

\section{Application to SIDM halos in simulations}
\label{sec:simulation_comparison}

\subsection{Simulation data \& analysis}
\label{ssec:simulation_data}

In this section, we will assess the density profiles in the context of numerical simulations. Here, we use the high-resolution N-body simulations of isolated SIDM halos evolved under velocity-independent cross sections presented in~\citep{Tran2024} and~\citep{Tran2025}. Each halo contains $3 \times 10^7$ DM particles within the virial radius ($r_{200}$) and is initialized with the NFW configuration as detailed in~\citep{Tran2024}. More detailed information of the simulations is provided in Appendix \ref{apd:sim_config}. 

The particle counts within each simulation snapshot are measured using 100 log-spaced radial bins from $0.01 \, r_{\rm{s}}$ to $3 \, r_{200}$, i.e. from around the value of the gravitational softening length to the largest sampling radius of the initial conditions at snapshot 0. The data are then further processed by merging shells with low particle counts and keeping only those within the virial radius. This is to avoid the strong exponential cut-off for $r > r_{200}$ configured in the initial conditions~\citep{Tran2024}. We limit the number of DM particles per shell to at least 400, ensuring a minimum signal-noise ratio (SNR) of $20$.

To directly measure the core density $\rho_{\rm{core}}$ and velocity dispersion $\sigma_{\rm{core}}$ for each snapshot, we start from the first radial bin $i$ with the cumulative particle count of $N_i \geq N_{\rm{min}}$. For the core density, we choose $N_{\rm{min}} = 10^3$, while for the core velocity dispersion, $N_{\rm{min}} = 10^4$. These choices are empirical and depend primarily on the simulation resolution. From the radial bin $i$, the cumulative average density $\bar\rho_i \pm \Delta{\bar\rho_i}$ is calculated. Comparing this to the density of the following radial bin $\rho_{i+1} \pm \Delta \rho_{i+1}$, if $\lvert \bar\rho_i - \rho_{i+1} \rvert \, \geq \left(\Delta{\bar\rho_i^2}+\Delta{\rho_{i+1}^2}\right)^{1/2}$, take $\bar \rho_i$ as the core density $\rho_{\rm{core}}$, else, consider the radial bin $i+1$ and repeat. For the core velocity dispersion $\sigma_{\rm{core}}$, we repeat the same process using the one-dimensional velocity dispersion profile $\sigma_{\rm{1D}} (r)$. For improved statistics, we also utilize a Monte Carlo (MC) approach, perturbing the density and one-dimensional velocity dispersion profiles according to their uncertainties, generating a set of samples. We then repeat the aforementioned process for the values of $\rho_{\rm{core}}$ and $\sigma_{\rm{core}}$ in each sample. The final results, derived from the distributions of sampled values, are consistent with those obtained directly from the original simulation configurations. The core half-density radius $r_{\rho/2}$ is obtained using a similar MC method.

To fit these sampled density structures using the analytical profiles detailed in Section \ref{sec:profiles}, we utilize the least-squares method in log space, minimizing the residual sum of squares (RSS)
\begin{equation}
    \label{eqn:chisqr}
    \chi^2 (\boldsymbol{\theta}) \propto \sum_i \Big( \log{\rho_i} - \log{\hat{\rho}_i (\rho_{\rm{c}}, r_{\rm{c}}, r_{\rm{s}}^{\prime}, n)} \Big)^2.
\end{equation}
Here, $\rho_i$ and $\hat{\rho}_i (\boldsymbol{\theta})$ are the density within radial bin $i$ and its expected value calculated from the analytical profile. The approach here is equivalent to taking the maximum likelihood approach, assuming the SNR of the density measurements remains constant across radial bins. This is typically untrue, as the SNR of the density measurement depends on the number of particles within each radial bin, which can vary by orders of magnitude. However, we want to prioritize the fitting of the halo core and can accept a reasonably good description of the density profile (with the enforced minimum SNR of $20$) rather than a statistically exact solution\footnote{Another choice of RSS can be found in~\citep{Fischer2024}, where the fitting criterion takes into account the Poisson statistics in isolated N-body simulations. This typically leads to the algorithm prioritizing fitting the outer region over the core, which is not ideal for the investigation of core collapse. The deviations of the results between the two choices of RSS are not significant, except in the values of $r_{\rm{s}}^\prime$ or when the halos are deep into core collapse at $T \gtrsim 0.8\,\tau$.}. To account for the statistical fluctuation of the density measurements, we employ a similar MC method as detailed above, i.e., perturbing the density profile and analyzing the resulting parameter distributions.


\subsection{Fitting results}
\label{ssec:fitting_results}

\begin{figure*}
    \centering
    \includegraphics[width=0.99 \textwidth]{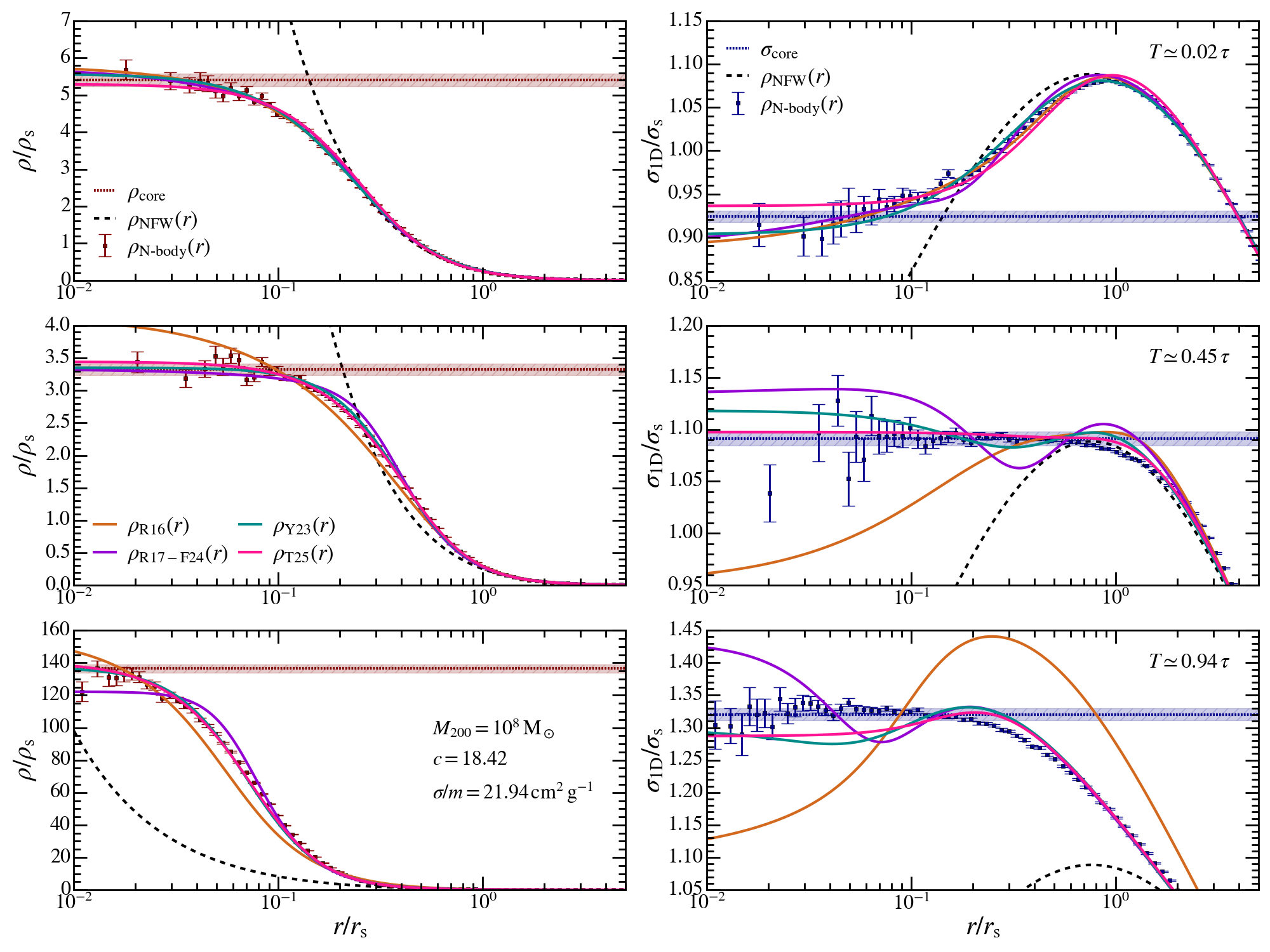}
    \caption{The density profile fits (left) and reconstructed one-dimensional velocity dispersion profiles (right) at three different snapshots for the halo of mass $10^8\msun$ evolved under the cross section of $21.94\cpm$. The first snapshot, $T \simeq 0.02\,\tau$ (top), corresponds to the earliest phase of core formation. The second, $T \simeq 0.45\,\tau$ (middle), represents a typical snapshot during core collapse, while the final snapshot, $T \simeq 0.94\,\tau$ (bottom), shows the halo deep into core collapse, during the gravothermal catastrophe. The scatter points and associated error bars represent measurements obtained from N-body simulations. The uncertainties for the density profile are derived from Poisson statistics, while the uncertainties in the velocity dispersion are quantified as the standard error of the standard deviation of the one-dimensional particle velocities. The fitted and reconstructed profiles for the Read (orange), Robertson-Fischer (purple), Yang (cyan), and our (pink) profiles are shown in solid lines. The dashed lines represent the initial NFW configuration for the halo of interest. The horizontal red and blue lines (and corresponding shaded areas) display the directly measured core density $\rho_{\rm{core}}$ and velocity dispersion $\sigma_{\rm{core}}$, respectively.}
    \label{fig:main_halo_fit}
\end{figure*}

\begin{figure}
    \centering
    \includegraphics[width=0.49 \textwidth]{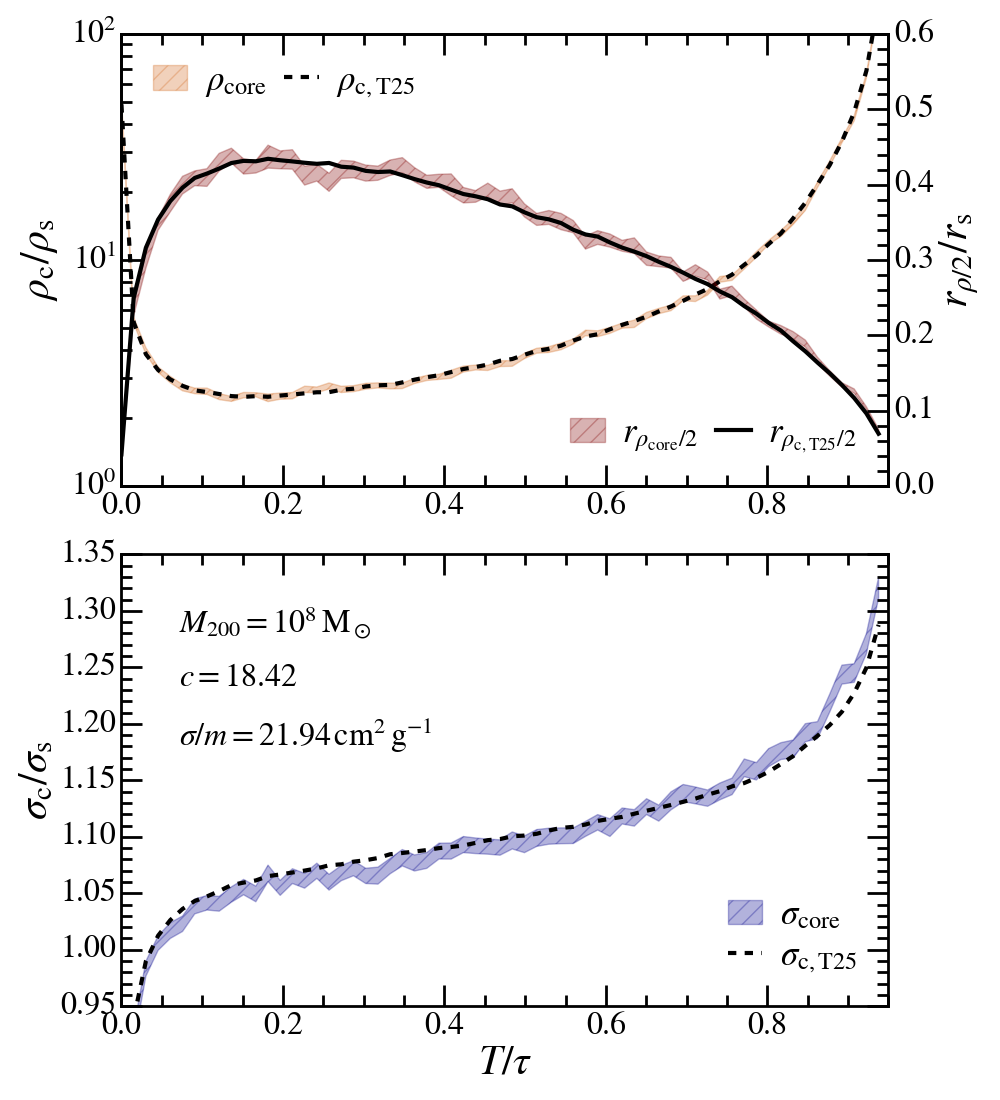}
    \caption{Comparisons between direct measurements (shaded areas) and fitted parameters using our density profile $\rho_{\rm{T25}}$ (black lines) for the core density $\rho_{\rm{c}}$ (top, left y-axis), core half-density radius $r_{\rho/2}$ (top, right y-axis), and core velocity dispersion $\sigma_{\rm{c}}$ (bottom) in the halo of mass $10^8\msun$ evolved under the cross section of $21.94\cpm$. The direct measurements are displayed with uncertainties, while the fitting errors of the parameters are negligible and therefore not shown.}
    \label{fig:params_evolution}
\end{figure}

\begin{figure}
    \centering
    \includegraphics[width=0.49 \textwidth]{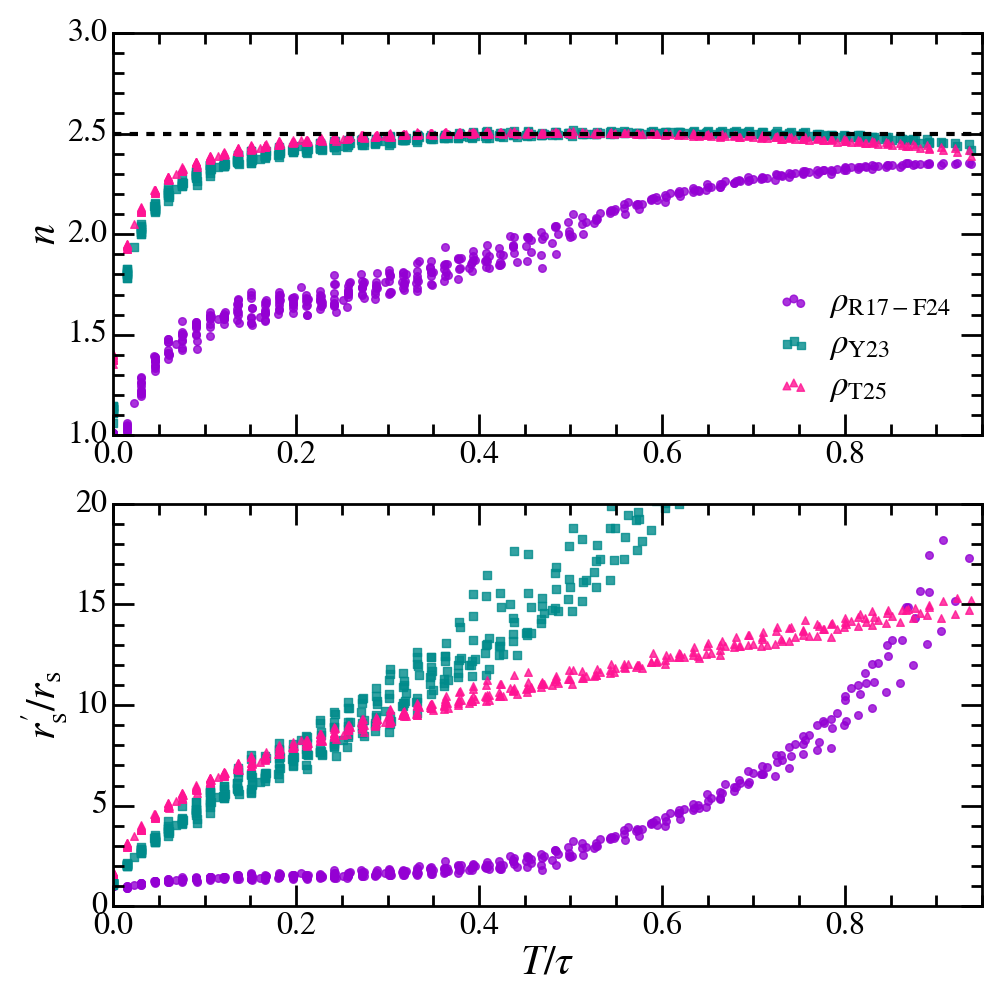}
    \caption{The evolutions of the fitted transition index $n$ (top) and scale radius $r_{\rm{s}}^{\prime}$ (bottom) for the Robertson-Fischer (purple), Yang (cyan) and our (pink) density profiles. Each data point corresponds to a single snapshot from an individual halo. All simulated halos are included in the figures. The dashed horizontal line represents a constant value of $n=2.5$, observed throughout most evolutionary stages in both the Yang and our profiles.}
    \label{fig:outer_region_param_evolution}
\end{figure}

For the following analysis, we focus on the halo of mass $10^8\msun$ evolved under the cross section of $21.94\cpm$. This system will be referred to as the main halo throughout the discussion. We scale all radii, densities, and velocity dispersions with $r_{\rm{s}}$, $\rho_{\rm{s}}$, and $\sigma_{\rm{s}} = \sqrt{G \, \rho_{\rm{s}} \, r_{\rm{s}}^2}$, while times are scaled with the collapse timescale $\tau$, defined in Equation 2.2 of~\citep{Yang2024} as
\begin{equation}
    \label{eqn:collapse_time_scale}
    \tau(\sigma_{\rm{eff}}/m) = \frac{150}{C} \frac{1}{\sigma_{\rm eff}/m} \frac{1}{\rho_{\rm{s}}} \left(\frac{1}{4 \pi \, \sigma_{\rm{s}}^2}\right)^{1/2}.
\end{equation}
In the current picture, $C = \mathcal{O}\left(1\right)$ often takes the value of $C \simeq 0.75$ to match the results of fluid models~\citep[e.g.][]{Essig2019,Yang2023,Zhong2023} and N-body simulations~\citep[e.g.][]{Koda2011,Essig2019,Shen2021,Yang2022}. Here, we use the value of $C \simeq 0.85$ in order to approximate $\tau$ as the onset of the gravothermal catastrophe, i.e. when the core density and velocity dispersion rapidly diverge while the core radius (and mass) vanishes.

Figure \ref{fig:main_halo_fit} shows the density profile fits and the reconstructed one-dimensional velocity dispersion profiles for three different snapshots of the main halo. The first snapshot, $T \simeq 0.02\,\tau$, corresponds to the earliest phase of core formation. The second, $T \simeq 0.45\,\tau$, represents a typical snapshot during core collapse, where self-similarity is most relevant. The results for this snapshot are representative of the halo's behavior within the time interval of $T \simeq 0.05 \text{--} 0.85 \, \tau$. The final snapshot, $T \simeq 0.94\,\tau$, shows the halo during the gravothermal catastrophe. The density profiles from each snapshot are fitted using the Read, Robertson--Fischer, Yang, and our profiles. Corresponding velocity dispersion profiles are then reconstructed based on the fitted parameters using the spherically symmetric Jeans equation (Equation~\ref{eqn:Jeans_equation}). For reference, the directly measured core density $\rho_{\rm{core}}$ and velocity dispersion $\sigma_{\rm{core}}$ are also displayed. We find that our profile and the Yang profile generally provide good fits to the halo density structures, while the Read profile deviates significantly after the earliest snapshots. This becomes more apparent in the logarithmic slope presented in Figure \ref{fig:log_slope}. Additionally, although the Robertson–Fischer profile provides a reasonable fit during the earlier evolutionary stages, it fails to accurately capture the density structure's behavior in the later phases of core collapse. The Yang profile and our profile yield notably similar results in fitting the density structures, with only minor deviations observed in the constant-density cores and the intermediate transition regions, i.e., the regions where density starts to decrease. However, these small deviations result in very different behaviors in the velocity dispersion profiles, most clearly seen in the middle panel of Figure \ref{fig:main_halo_fit}, with only our profile closely matching simulations and maintaining consistency with the isothermal-core configuration. It should be noted, nonetheless, that even our profile exhibits slight deviations from the simulated velocity dispersion around $r_{\rm{s}}$. The underlying reason can be inferred from the deviations in the logarithmic slope, which also indicate that achieving a more accurate reconstruction of the velocity dispersion profile would require adopting a more complex functional form with additional degrees of freedom. The exceptions for the general goodness of fit observation occur only at the earliest snapshots, where the Yang and other profiles provide a better description of the velocity dispersion structures, and at the later stages of core collapse, where none of the profiles can accurately capture the isothermal core.

Figure \ref{fig:params_evolution} compares the directly measured values of core density, half-density radius, and velocity dispersion with those inferred from fitting using our density profile. The direct measurements are shown with associated uncertainties, while the inferred values are displayed without. This is due to the negligible fitting errors, calculated using the bootstrapping approach detailed in Section \ref{ssec:simulation_data}. We observe that the values remain generally consistent, even during the early stages of core formation, and continue to hold up as the system approaches the gravothermal catastrophe. This is as expected with the performance of the density profile fits and the velocity dispersion reconstructions discussed above.


Given that both the Yang profile and our profile perform well in representing the density structure of halos across different evolutionary stages, resulting in similar values of $\rho_{\rm{c}}$ and $r_{\rho/2}$, it is of particular interest to examine the evolutions of the transition index $n$ and the scale radius $r_{\rm{s}}^{\prime}$ over time. Figure \ref{fig:outer_region_param_evolution} presents these evolutionary trends, along with the corresponding parameter evolution derived from the Robertson-Fischer profile. In both the Yang profile and our profile, we observe that the transition index $n$ evolves rapidly from an initial value of $n = 1$, characteristic of the NFW profile, to approximately $n \simeq 2.5$, where it remains relatively stable throughout the majority of the halo's evolution. As discussed in Section \ref{ssec:core_profiles}, this behavior is indicative of the preservation of self-similarity during the evolutionary process. This relative stability of $n$ is in contrast to the evolutions of the scale radius $r_{\rm{s}}^{\prime}$, which vary significantly across epochs. Such deviations may stem from limitations in the density profile fitting, where variations in $r_{\rm{s}}^{\prime}$ compensate for the need to simultaneously fit both the core and the outer regions of the halo. Nevertheless, our density profile continues to exhibit a consistent and coherent evolution of $r_{\rm{s}}^{\prime}$ across different halos. On the other hand, the Yang profile shows significant scatter in the evolution of $r_{\rm{s}}^{\prime}$, breaking self-similarity and displaying a lack of systematic behavior. Interestingly, unlike in the Yang or our profiles, the scale radius in the Robertson-Fischer profile retains the value of the initial NFW scale radius throughout a significant portion of the evolutionary sequence. A similar behavior is also observed in the Read profile, which, although not shown, preserves this initial scale radius throughout the entire gravothermal collapse process.

\section{Conclusions and Discussions}
\label{sec:conclusions}

In this work, we introduce a novel density profile for SIDM halos (Equation~\ref{eqn:TVS_density_profile}) that closely approximates the isothermality of the core. We show that this profile provides a good representation of DM halos in SIDM models, while retaining the simplicity of an analytic functional form. We find that
\begin{enumerate}
    \item Analytically, halos with structure following our density profile possess close-to-flat velocity dispersions extending to a few of the characteristic core radii (i.e. the typical size of the core region). This results in a close agreement with the isothermal-core configuration seen in the analytical self-similar solutions presented in~\citep{Balberg2002,Lynden-Bell+Eggleton1980}.
    \item The density structure fits performed with our density profile result in high levels of stability and the best density profile fits to various simulation data. The core density and half-density radius inferred from the fitted profiles show strong agreement with the direct measurements from simulations. Notably, the fitted values benefit from significantly lower uncertainties, estimated via statistical bootstrapping, compared to the uncertainties associated with direct measurements.
    \item The velocity dispersion profiles reconstructed from the density structure fits using our profile exhibit close agreement with simulation data, whereas those derived from other common functional forms fail to adequately capture the isothermal-core configuration.
\end{enumerate}

With the exception of the earliest stages of core formation and the late-time deep core collapse regime, our profile demonstrates a high level of agreement with simulation data. As a result of its ability to accurately represent dark matter halos in self-similar scenarios, our profile serves as a promising tool for reducing reliance on computationally expensive SIDM simulations of velocity-independent cross section (VICS) models. The profile can act as a benchmark for comparisons between VICS and more complex interaction models using the empirical formulae detailed in Appendix \ref{apd:empirical_results}. The resulting values of the transition index $n$, derived from the density profile fits, may also serve as a diagnostic for self-similarity in more intricate SIDM models. Additionally, having a functional form also facilitates accurate and efficient calculations of the heat conductivity and luminosity structures of DM halos (an example of which is shown in Appendix \ref{apd:luminosity}). Finally, initial conditions for simulations entering the deep core collapse regime may be constructed directly from our density profile, following the Eddington sampling procedure described in~\citep{Tran2024}, thus mitigating the need to evolve halos from earlier epochs.

\begin{acknowledgments}
We thank the anonymous referee for useful comments and suggestions. We thank Philip Harris for useful discussions.
\end{acknowledgments}

\bibliography{main.bib}

\appendix

\section{Analytical calculation of the core velocity dispersion via the isothermal K-G ratio}
\label{apd:sigma_c_T25}

To inspect the core isothermal K-G ratio (Equation \ref{eqn:isothermal_K-G_ratio}) of our density profile, we first separate $\bar{\rho}_{\rm{T25}}\left(r\right)$ into
\begin{align}
    \label{eqn:inner_density_profile}
    g(\bar{r}) &= \left(\frac{\tanh{\bar{r}}}{\bar{r}}\right)^n, \\
    \label{eqn:density_profile_edge_modifier}
    h(\bar{r}) &= \frac{1}{\left(1 + \left(\lambda \bar{r}\right)^\gamma\right)^{\frac{3-n}{\gamma}}},
\end{align}
with $\lambda = r_{\rm{c}} / r_{\rm{s}}^{\prime}$. Here, $g(\bar{r})$ and $h(\bar{r})$ represent the approximate density profile of the halo inner part and the modifier for the halo outer edge. Taking the $\bar{r} \ll 1$ approximations, we obtain the first and second derivatives
\begin{align}
    \label{eqn:g_prime_approx}
    g^{\prime} (\bar{r}) &= - \frac{2 n}{3} \bar{r} + \mathcal{O}(\bar{r}^3), \\
    \label{eqn:g_prime_prime_approx}
    g^{\prime\prime} (\bar{r}) &= - \frac{2 n}{3} + \frac{4n^2}{9} \bar{r}^2 + \mathcal{O}(\bar{r}^3), \\
    \label{eqn:h_prime_approx}
    h^{\prime} (\bar{r}) &= - \left(3-n\right) \frac{\left(\lambda \bar{r}\right)^\gamma}{\bar{r}} + \mathcal{O}(\bar{r}^{2\gamma-1}), \\
    \label{eqn:h_prime_prime_approx}
    h^{\prime\prime} (\bar{r}) &= - \left(3-n\right) \left(\gamma - 1\right) \frac{\left(\lambda \bar{r}\right)^\gamma}{\bar{r}^2} + \mathcal{O}(\bar{r}^{2\gamma-2}).
\end{align}

Considering only the simplified trial density profile, $\rho (\bar{r}) = g (\bar{r})$, the resulting core isothermal K-G ratio takes the form of
\begin{equation}
    \label{eqn:TVS_simple_isothermal_K-G_ratio}
    I^{\ast}_{\rm{c,T25}} = \frac{1}{2n} + \mathcal{O}\left(\bar{r}^2\right),
\end{equation}
which is independent of radius (to the order of $\bar{r}^2$). 
When the outer edge modifier $h (\bar{r})$ is included, the core isothermal K-G ratio would follow
\begin{equation}
    \label{eqn:TVS_isothermal_K-G_ratio}
    I_{\rm{c,T25}} \approx \left( 2n + \left(3-n\right)\left(\gamma+1\right)\frac{\left(\lambda \bar{r}\right)^\gamma}{\bar{r}^2} \right)^{-1}.
\end{equation}
Here, the extra term $- \left(3-n\right)^2 \left(\lambda \bar{r}\right)^{2 \gamma} / \bar{r}^2$ inside the parentheses has been ignored. It is clear from Equation \ref{eqn:TVS_isothermal_K-G_ratio} that as $\bar{r} \rightarrow 0$, unless $\gamma \geq 2$, $\left(\lambda \bar{r}\right)^{\gamma} / \bar{r}^2 \rightarrow \infty$ and $I_{\rm{c,T25}} \rightarrow 0$. In fact, $\gamma = 2$ represents a special case with
\begin{equation}
    \label{eqn:TVS_isothermal_K-G_ratio_gamma_2}
    I_{\rm{c,T25}} =  \frac{1}{2n + 3\left(3-n\right) \lambda^2} + \mathcal{O}\left(\bar{r}^2\right),
\end{equation}
while, otherwise,
\begin{equation}
    \label{eqn:TVS_isothermal_K-G_ratio_gamma_not_2}
    I_{\rm{c,T25}} = \frac{1}{2n + \left(3-n\right)\left(\gamma+1\right) \lambda^{\gamma} \, \bar{r}^{\gamma-2}} + \mathcal{O}\left(\bar{r}^2\right).
\end{equation}
The first case provides a near-optimal situation, with $I_{\rm{c,T25}}$ becoming practically independent of $\bar{r}$ similar to in the idealistic configuration. In most cases, $\lambda \ll 1$, as well as $n$ taking values close to 3, resulting in the core isothermal K-G ratio reverting to the form displayed by Equation \ref{eqn:TVS_simple_isothermal_K-G_ratio}.

\section{Simulation configurations}
\label{apd:sim_config}

\begin{table}[h]
    \centering
    \addtolength{\tabcolsep}{2.5pt}
    \def\arraystretch{1.5}
    \begin{tabular}{c c c c c c c c c c}
        \hline
        $\log M_{200}$ & $m_{\rm{DM}}$ & $c$ & $r_{\rm s}$ & $\log \rho_{\rm s}$ & $\sigma/m$ \\ [0.25ex] 
        [${\rm M}_\odot$] & [${\rm M}_\odot$] &  & [$\rm kpc$] & [${\rm M}_\odot\,{\rm kpc}^{-3}$ ] & [${\rm cm}^2\,{\rm g}^{-1}$] \\ [1ex] 
        \hline\hline

        7.9 & 2.64 & 18.70 & 0.49 & 7.43 & 31.98\\
        \hline
        
        7 & 0.33 & 21.21 & 0.21 & 7.57 & 24.94, 19.81\\ 
 
        7.5 & 1.05 & 19.81 & 0.34 & 7.50 & 43.54, 19.76\\

        8 & 3.33 & 18.42 & 0.53 & 7.42 & 28.07, 21.94\\

        8.5 & 10.5 & 17.05 & 0.84 & 7.33 & 18.60, 14.62\\
        
        9 & 33.3 & 15.69 & 1.35 & 7.24 & 13.09, 8.87\\ 
        \hline
    \end{tabular}
    \caption{Simulation configurations of the smaller data set (top)~\citep{Tran2024} and larger data set (bottom)~\citep{Tran2025}. (1) $M_{200}$ and (2) $m_{\rm{DM}}$ are the virial mass of the halo and the mass of DM particles in the simulation, respectively. (3) $c$ is the halo concentration parameter. (5) $\rho_{\rm s}$ and (4) $r_{\rm s}$ are the scale density and radius of the NFW profile. (6) $\sigma/m$ is the SIDM collisions' cross section per unit of mass.}
    \label{tab:halo_configurations}
\end{table}

Table \ref{tab:halo_configurations} shows the detailed simulation configurations in the two data sets~\citep{Tran2024,Tran2025}. For the smaller data set~\citep{Tran2024}, two velocity-independent cross sections for DM self-interactions are presented in the original work; however, we focus mainly on the lower value of $\sigma/m = 31.98 \cpm$ to avoid numerical effects causing deviation in the self-similarity~\citep{Mace24, Palubski2024, Shen2024}. For the larger data set~\citep{Tran2025}, we utilize the results of all simulations with velocity-independent cross sections.

\section{Logarithmic slope of density profiles}
\label{apd:log_slope}

\begin{figure}
    \centering
    \includegraphics[width=0.49 \textwidth]{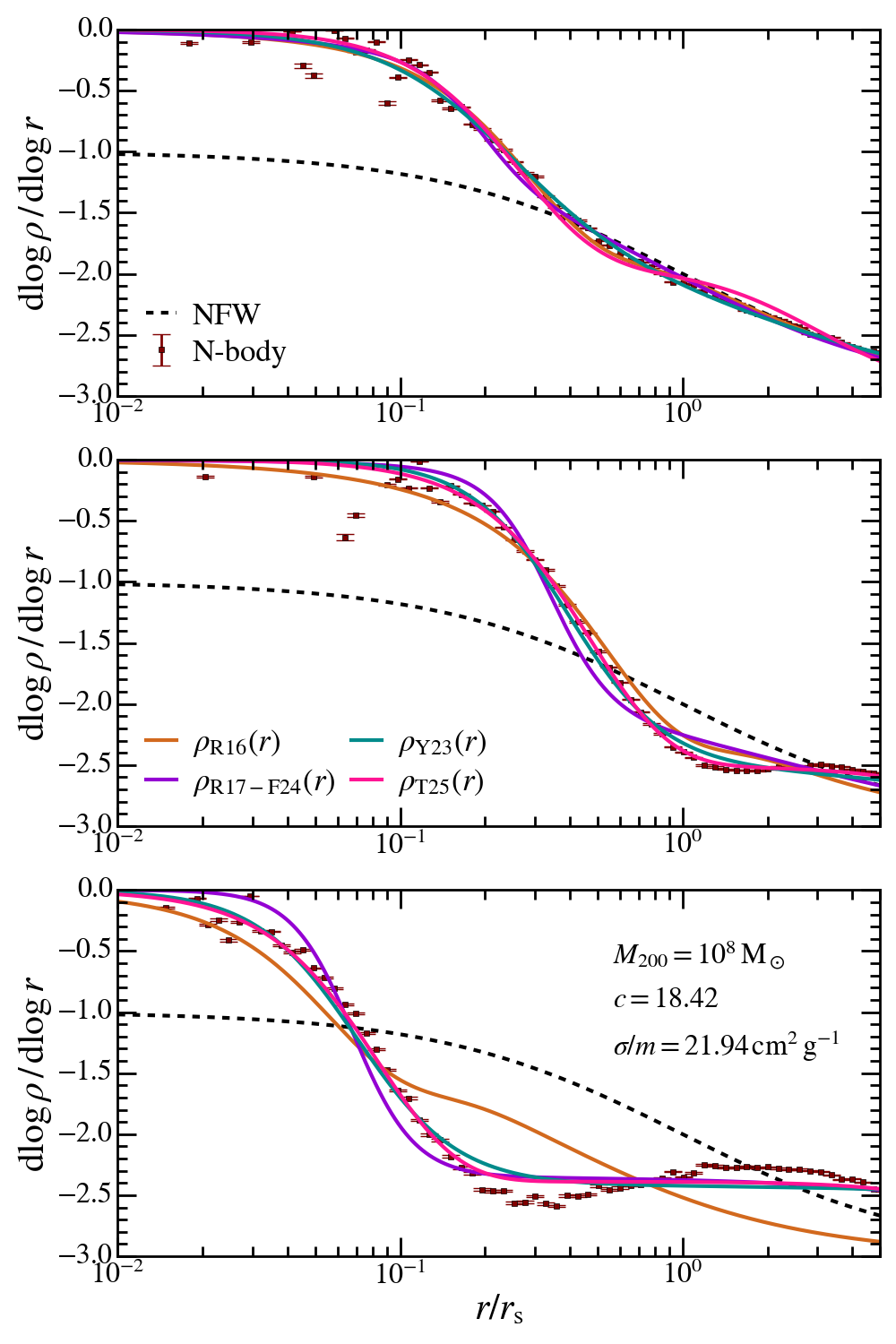}
    \caption{The logarithmic slopes of the density profile fits at three different snapshots for the halo of mass $10^8\msun$ evolved under the cross section of $21.94\cpm$. The snapshots and halo are the same as in Figure \ref{fig:main_halo_fit}. The scatter points and associated error bars represent measurements obtained from N-body simulations. The profiles for the Read (orange), Robertson-Fischer (purple), Yang (cyan), and our (pink) profiles are shown in solid lines. The dashed lines represent the initial NFW configuration for the halo of interest.}
    \label{fig:log_slope}
\end{figure}

Figure \ref{fig:log_slope} presents the logarithmic slopes of the fitted density profiles shown in Figure \ref{fig:main_halo_fit}. The level of agreement between the logarithmic slopes of the fitted profiles and those from the simulations reveals why, despite the density profiles being well reproduced by our functional form, the reconstructed velocity dispersion profiles still exhibit small deviations. Nevertheless, our profile remains the closest match to the log–log slope of the density structure. In order to achieve a better fit to the slope and a more accurate reconstruction of the velocity dispersion profiles, a more complex functional form with additional degrees of freedom is required.

\section{Empirical evolutions of parameters in the self-similar solution}
\label{apd:empirical_results}

Using the fitting results of Section \ref{ssec:fitting_results}, we arrive at a general set of empirical equations describing the evolutions of $\rho_{c,T25}$, $r_{\rm{c,T25}}$, $r_{\rm{s,T25}}$, and $n_{\rm{T25}}$ in term of the scaled time $t = T/\tau$ (Equation \ref{eqn:collapse_time_scale}). These equations can serve as good references when comparisons between simple VICS models and more complicated SIDM models are needed, especially in the context of idealized isolated halos. For simplicity, we construct the empirical equations only out of polynomials and square/cube roots. Constraints in the initial conditions are also enforced to map the density profile back to the NFW configuration. These include $r_{\rm{s},t=0}^{\prime} = r_{\rm{s}}$ and $n_{t=0} = 1$. Idealistically, the condition of $\rho_{\rm{c},t=0} \, r_{\rm{c},t=0} = \rho_{\rm{s}} \, r_{\rm{s}}$ should also be satisfied. However, we find that this is a difficult constraint to implement and, thus, ignore it. The resulting empirical fits then follow

\begin{multline}
   \label{eqn:rho_c_empirical_fit}
   \log{\left(\frac{\rho_{\rm{c,T25}}}{\rho_{\rm{s}}}\right)} = \frac{1}{1 - t} \bigg( 0.592 - 25.71 \, t + 13.02 \, t^2 \\
   - 4.426 \, t^3 + 24.22 \, \sqrt{t} - 7.363 \, \sqrt[3]{t} \, \bigg)^{-1},
\end{multline}
\begin{multline}
   \label{eqn:r_c_empirical_fit}
   \hspace{1.1cm} \frac{r_{\rm{c,T25}}}{r_{\rm{s}}} = - 3.110 \, t + 2.066 \, t^2 - 1.149 \, t^3 \\
   + 2.229 \, \sqrt{t} - 0.044 \, \sqrt[3]{t},
\end{multline}
\begin{multline}
    \label{eqn:r_s_empirical_fit}
    \hspace{1.1cm} \frac{r_{\rm{s,T25}}}{r_{\rm{s}}} = 1 + 12.455 \, t - 8.785 \, t^2 + 0.556 \, t^3 \\
    - 8.364 \, \sqrt{t} + 14.706 \, \sqrt[3]{t},
\end{multline}
\begin{multline}
\label{eqn:n_empirical_fit}
    \hspace{1.35cm} n_{\rm{T25}} = 1 - 1.978 \, t + 2.417 \, t^2 - 1.293 \, t^3 \\
    - 3.550 \, \sqrt{t} + 5.745 \, \sqrt[3]{t}. 
\end{multline}

\section{Luminosity \& heat conductivity}
\label{apd:luminosity}

\begin{figure}
    \centering
    \includegraphics[width=0.49 \textwidth]{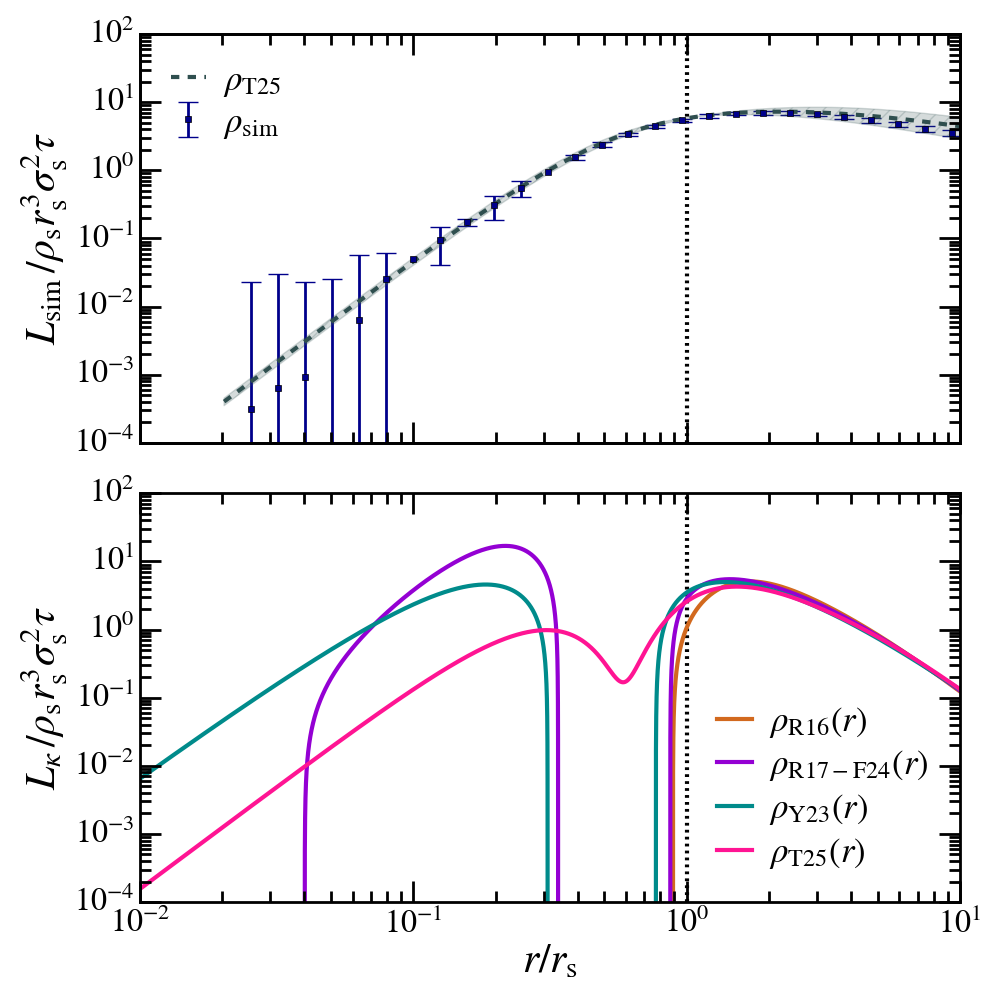}
    \caption{The luminosity profiles of the $10^8\msun$ halo evolved with $\sigma/m = 21.94\cpm$, calculated at $T \simeq 0.45 \, \tau$. Top: The scatter points and dashed lines are derived from the time evolution of the halo, based on interpolation of the simulated profiles and on the corresponding fitted density and reconstructed velocity dispersion using our density model, respectively. Bottom: The solid lines show the luminosity profiles computed from the velocity dispersion gradient of the four fitted density profiles and the empirical heat conductivity formula (Equation \ref{eqn:kappa}) with $\beta = 0.75$.}
    \label{fig:luminosity}
\end{figure}


To illustrate a potential application of our density profile, we compute the luminosity profile of a halo during the core-collapse phase following the same procedure as in~\citep{Yang2022}, i.e.
\begin{equation}
    \label{eqn:luminousity}
    L(r) = -4\pi\int_0^r \mathrm{d}r'\, {r'}^2\,\tilde{\rho}(r')\,\frac{\mathrm{D}E(r')}{\mathrm{D}T},
\end{equation}
where the Lagrangian time derivative of the specific energy is estimated by
\begin{equation}
    \label{eqn:lagrangian_energy_derivative}
    \frac{\mathrm{D}E(r)}{\mathrm{D}T} = \frac{E(r_M,T+\Delta T) - E(r,T)}{\Delta T}.
\end{equation}
Here, $E(r) = 3\sigma_{\rm{1D}}^2(r)/2+ \Phi(r)$, and $r_M$ is the radius that encloses the same mass at time $T + \Delta T$ as the enclosed mass within $r$ at time $T$. In practice, we calculate $\mathrm{D}E/\mathrm{D}T$ as the symmetric difference averaged over $T-\Delta T$ and $T+\Delta T$ with $\Delta T\simeq 0.015 \, \tau$. Figure \ref{fig:luminosity} shows these profiles, calculated using values of $\sigma_{\rm{1D}}$ and $\Psi$ interpolated from simulation data, as well as those reconstructed from the fitted density profile.

Additionally, we adopt an alternative approach in which the luminosity is computed directly from the fitted analytical profile
\begin{equation}
    \label{eqn:luminosity_kappa}
    L(r) = - 4\pi r^2 \kappa \frac{{\rm{d}}\sigma_{\rm{1D}}^2}{{\rm{d}}r},
\end{equation}
with
\begin{equation}
    \label{eqn:kappa}
    \kappa = 0.27 \beta \, \rho \, \sigma_{\rm{1D}}^3 \frac{\sigma / m}{G},
\end{equation}
where typically $\beta = 0.75$ for isolated halos~\citep{Essig2019}. $\rho$ and $\sigma_{\rm{1D}}$ are taken from the density fits and the velocity dispersion reconstructions of the four density profile choices. The results of such calculations are also shown in Figure \ref{fig:luminosity}. Although the approaches described by Equations \ref{eqn:luminousity} and \ref{eqn:luminosity_kappa} do not necessarily converge, in the inner regions of the halo, the time evolution–based calculation and the reconstructed velocity dispersion gradient approach based on our fitted profile yield fairly similar luminosity profiles.

\end{document}